# Time Scaling Relations for Step Bunches from Models with Step-Step Attractions (B1-Type Models)

A. Krasteva[1, a)], N. Akutsu[2, b)] and V. Tonchev[3, c)]

[1]*Institute of Electronics, BAS, 72 Tzarigradsko chaussee blvd, 1784 Sofia, Bulgaria*
[2]*Osaka Electro-Communication University, Neyagawa, Osaka 572-8530, Japan*
[3]*Institute of Physical Chemistry, BAS, Acad. G. Bonchev Str., block 11, 1113 Sofia, Bulgaria*

[a)] Corresponding author: anna0kr0stz@gmail.com
[b)] nori3@phys.osakac.ac.jp
[c)] tonchev@ipc.bas.bg

**Abstract.** The step bunching instability is studied in three models of step motion defined in terms of ordinary differential equations (ODE). The source of instability in these models is step-step attraction, it is opposed by step-step repulsion and the developing surface patterns reflect the balance between the two. The first model, TE2, is a generalization of the seminal model of Tersoff et al. (1995). The second one, LW2, is obtained from the model of Liu and Weeks (1998) using the repulsions term to construct the attractions one with retained possibility to change the parameters in the two independently. The third model, MM2, is a minimal one constructed *ad hoc* and in this article it plays a central role. New scheme for scaling the ODE in vicinal studies is applied towards deciphering the pre-factors in the time-scaling relations. In all these models the patterned surface is self-similar - only one length scale is necessary to describe its evolution (hence B1-type). The bunches form finite angles with the terraces. Integrating numerically the equations for step motion and changing systematically the parameters we obtain the overall dependence of time-scaling exponent $\beta$ on the power of step-step attractions $p$ as $\beta = 1/(3+p)$ for MM2 and hypothesize based on restricted set of data that it is $\beta = 1/(5+p)$ for LW2 and TE2.

## INTRODUCTION

Surface patterning is at present subject of increasing interest due to the significant progress in the instrumentation operating easily on the atomic scale. Another reason is the practical accessibility of the patterned surfaces as nano-templates [1, 2]. These developments require development of adequate models [3] but also further detail of the theories from the past decades. Here we revisit a theory of step bunching due to step-step attraction that is introduced in order to study the consequences of vicinal hetero-epitaxial growth [4] in the course of which the vicinal surface becomes strained due to a lattice misfit. The theory is based on the formulation and numerical solution of a system of ordinary differential equations that are highly non-linear. In achieving our aim we take into account also recent developments [5, 6, 7] that remain aside the attention of the community so far.

Here we solve a particular problem – when using a general form of the step-step attractions with a generalizing exponent $p$ (the range of attraction) how is it reflected in the time-scaling exponent $\beta$ in the time-scaling of the bunch size that quantifies the bunching phenomenon.

## THE MODELS

We start [5] with a rather general equation that describes the motion of the $i$-th step from the vicinal stairway:

$$\frac{dx_i}{dt} = ka_i + ur_i \qquad (1)$$

where $a_i$ and $r_i$ (different for each equation) are written in terms of distances between nearest steps only $\Delta x_i \equiv x_i - x_{i-1}$ and have dimensions of $[a_i] = L(ength)^q$, $[r_i] = L^m$. Since the dimension of $dx_i/dt$ is $LT(ime)^{-1}$ the resulting dimensions of $k$ and $r$ (uniform for all equations) are $L^{1-q} T^{-1}$ and $L^{1-m} T^{-1}$ respectively. The system of equations (1) is defined together with the initial condition $\Delta x_i^0 \equiv l_0$ for each value of $i$, manifesting the structural coherence of the vicinal crystal. For non-dimensionalization we use time- and length-scales $\tau$ and $\xi$ to obtain:

$$\frac{dX_i}{dT} = \xi^{q-1} k \tau A_i + \xi^{m-1} \tau u R_i \tag{2}$$

with the dimensionless quantities $A_i$, $R_i$, $X_i \equiv x_i \xi^{-1}$ and $T \equiv t \tau^{-1}$. The initial condition is now $L_0 \equiv l_0 \xi^{-1}$. The length- and time-scales are still not specified and they could be set such that:

$$\xi^{q-1} k \tau = 1; \quad \xi^{m-1} \tau u = 1 \tag{3}$$

and from the two equations for two variables we obtain:

$$\tau = \left(ku^{-1}\right)^{\frac{1-q}{m-q}} k^{-1}; \quad \xi = \left(ku^{-1}\right)^{\frac{1}{m-q}} \tag{4}$$

This procedure leaves the system, eq. (1), without parameters except $L_0$:

$$\frac{dX_i}{dT} = A_i + R_i \tag{5}$$

Now if we look for self-similar solution to eq. (1) for a characteristic length $w$ of the form $wt^\delta = s_{wt}$ and the scaling parameter $s_{wt}$ and the scaling exponent $\delta$ are unknown, the solution of eq. (5) preserves the scaling exponent $\delta$ and could be given in the form $WT^\delta = S_{WT}$ containing a different from $s_{wt}$ scaling parameter $S_{WT}$. Now, we can obtain a link between the two scaling parameters by noting that $W \equiv w/\xi$ and $T \equiv t/\tau$: $WT^\delta = \frac{w}{\xi}\left(\frac{t}{\tau}\right)^\delta = S_{WT}$ and thus to obtain the originally sought parameter $s_{wt}$:

$$wt^\delta = \left(ku^{-1}\right)^{\frac{1}{m-q}} \left[\left(ku^{-1}\right)^{\frac{1-q}{m-q}} k^{-1}\right]^\delta S_{WT} = s_{wt} \tag{6}$$

The effect of eq. (6) is threefold – (i) part of the road in finding $s_{wt}$ is passed by a straightforward dimensional analysis; (ii) the other part, the determination of $S_{WT}$ is done using the scaled eq. (5) which contains only one parameter, $L_0$, instead of 3 - $k$, $u$ and $l_0$, and since normally runs with 10 different values of any parameter are needed, the reduction of computational time is 100 times, note that, by looking at eq. (6), it is hard to imagine that even 1000 calculations with eq. (1) would result in obtaining $s_{wt}$; (iii) in each calculation of the derivative in eq. (5) two less multiplications are done in comparison with eq.(1). In the table below are presented the three models under consideration.

|  | $A_i$ | $R_i$ |
|---|---|---|
| MM2 | $-F_i^{-(p+1)} \equiv -\left(\Delta X_i^{-(p+1)} - \Delta X_{i+1}^{-(p+1)}\right)$ | $F_i^{-(n+1)} \equiv \Delta X_i^{-(n+1)} - \Delta X_{i+1}^{-(n+1)}$ |
| LW2 | $-\left(2F_i^{-(p+1)} - F_{i+1}^{-(p+1)} - F_{i-1}^{-(p+1)}\right)$ | $2F_i^{-(n+1)} - F_{i+1}^{-(n+1)} - F_{i-1}^{-(n+1)}$ |
| TE2 | $-\left[\Delta X_{i+1}^{-1}\left(F_{i+2}^{-(p+1)} - F_{i+1}^{-(p+1)}\right) - \Delta X_i^{-1}\left(F_{i+1}^{-(n+1)} - F_i^{-(n+1)}\right)\right]$ | $\Delta X_{i+1}^{-1}\left(F_{i+2}^{-(n+1)} - F_{i+1}^{-(n+1)}\right) - \Delta X_i^{-1}\left(F_{i+1}^{-(n+1)} - F_i^{-(n+1)}\right)$ |

## NUMERICAL RESULTS

Here are presented qualitative data on the surface profiles, surface slopes and step trajectories and then, in the second part, quantitative data on time evolution of the studied surfaces. During the numerical calculations we use the monitoring protocol introduced in [8] in order to obtain quantitative data for the monitored surfaces.

## Qualitative Results – Step Trajectories and Surface Profiles

In Figure 1 are shown qualitative data on the surface profiles and step trajectories for the case of TE2. These results are collected for $n = 2$ and $p = 0$. It is clearly seen that when two bunches collide (fig.1 a) the width of the new bunch is equal to the sum of the widths of the two colliding bunches. In fig. 1 b is shown the surface profile and in fig. 1c is shown the surface slope. It is seen that the slope remains constant although the width of the "peaks" is different and equal to the bunch width.

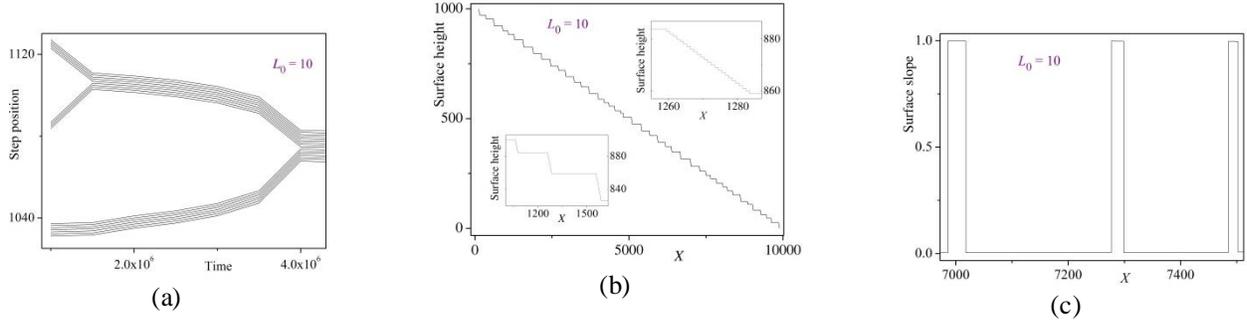

**FIGURE 1**. Trajectories (a), surface profiles at different magnifications (b) and surface slope (c) from TE2 with $n = 2$ and $p = 0$.

## Quantitative Results for the Surface Time Evolution

In Figure 2 are presented data from TE2 for the dependence on the bunch size of the minimal distance in the bunch (fig 2a) and of the bunch width (fig 2a, inset) while in fig2b is presented the time evolution of the bunch size $N$. The time scaling exponent of $N$ is always $1/5$ when $p = 0$ different from the finding of Tersoff et al. [4] $\beta$ 0.25 but their data spans values of $N$ up to 10 which is not enough to obtain a true asymptotics.

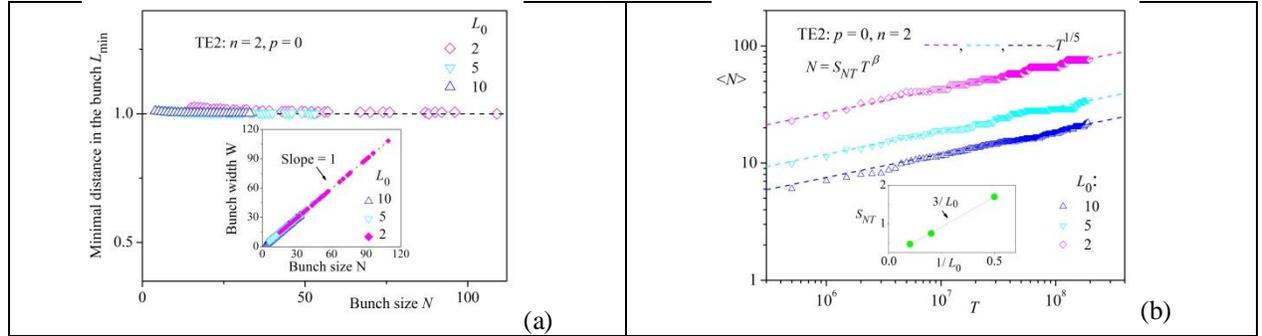

**FIGURE 2**. Figure 2a shows the dependence of the minimal distance in the bunch on the bunch size, it is constant, and in the inset is shown the dependence of the bunch width on the bunch size, it is linear with slope 1 for the three models but shown for TE2 only, hence B1-type [9], note the difference with [10, 11]. Figure 2b shows the time-dependence of the bunch size, $\beta(p=0) = 1/5$, the scaling prefactor is $3/L0$ (inset). We observe no dependence of $\beta$ on $n$ for fixed values of $p$.

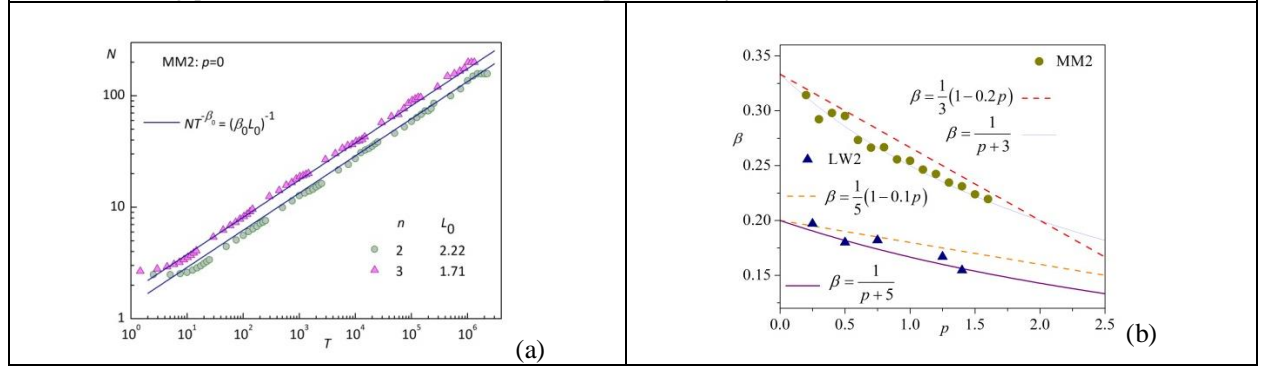

**FIGURE 3**. Time-scaling of (a) the average bunch size *N* from MM2 for *p* = 0, $\beta_0$ = 1/3; (b) dependence of *β* on *p*, it is well studied in MM2 since the starting point for the decrease with increasing *p* is $\beta_0$ = 1/3, thus for *p* = 2 would go down to 1/5 changing with 0.13333 while for the LW2 model $\beta_0$ = 1/5 and for the range of *p* = 0 ÷ 2 *β* would change only with 0.04÷0.06.

In Figure 3a are presented data for the time evolution of the bunch size *N* from MM2 for *p* = 0 and the time-scaling exponent is 1/3 different from the case of TE2. Different values of *β* are collected on Fig 3b as function of the exponent *p* describing the range of attraction, when *p* = 0 it is of infinite range. The calculations with MM2 are quite easier – in order to have ~100 steps in average in a bunch one should go to times of $10^6$ in the best case of *p* = 0 while the same size is achieved with TE2 and LW2 with times of $10^{10}$.

## CONCLUSION

We studied the step bunching process in three models where the source of the instability is step-step attraction. Our most important finding answers still unanswered question – what is the dependence of the time-scaling exponent *β* when changing the values of the exponents *n* and *p* that describe the range of step-step repulsion and attraction, respectively. There is no dependence on *n* and the dependence on *p* is *β*=1/(*p*+3) for MM2. For the other two models, TE2 and LW2, we hypothesize that it is *β*=1/(*p*+5) mainly in analogy with MM2 but only further study could prove this.

## ACKNOWLEDGMENTS

This work is supported by Bulgarian NSF, grant No. T02-8/121214, and Japanese Society for the Promotion of Science (JSPS) KAKENHI, grant No. 25400413.